\documentclass[10pt]{amsart}
\usepackage{amssymb}
\usepackage{enumerate}
\begin{document}
\begin{abstract}
Using the theory of optimal rocket trajectories in general relativity, recently developed in \cite{HN11}, we show that the ``obvious'' manoeuvre of using a tangential instantaneous acceleration to escape a stable circular orbit in the Schwarzschild spacetime satisfies the optimality conditions if and only if the magnitude of the acceleration is smaller than a certain bound. 
\end{abstract}
%
%
\title{Optimal escape from circular orbits around black holes}
\author{Jos\'{e} Nat\'{a}rio}
\address{Centro de An\'alise Matem\'atica, Geometria e Sistemas Din\^amicos, Departamento de Matem\'atica, Instituto Superior T\'ecnico, 1049-001 Lisboa, Portugal}
\thanks{Partially supported by FCT (Portugal).}
\maketitle
%
%
%
\section*{Introduction}
Using Newtonian mechanics, the energy per unit mass acquired by a rocket on an instantaneous acceleration which changes its velocity from $\vec{v}$ to $\vec{v} + \Delta \vec{v}$ is
\[
\Delta E = \frac12 \left|\vec{v} + \Delta \vec{v}\,\right|^2 - \frac12 |\vec{v}\,|^2 = \frac12 |{\Delta\vec{v}}\,|^2 + \langle\vec{v},\Delta\vec{v}\,\rangle.
\]
Therefore, the energy is maximized when the instantaneous acceleration is parallel to the initial velocity.

Now assume that the rocket is on a circular orbit of radius $r$ around a spherically symmetric body of mass $M$. Its initial speed is then\footnote{We use the conventions of \cite{MTW73}, including geometrized units ($c=G=1$).}
\[
V = \left(\frac{M}{r}\right)^\frac12.
\]
An obvious strategy to escape the gravitational field of the mass $M$ is to accelerate instantaneously in the tangential direction, thus maximizing the energy per unit mass. However, as first noted by Lawden \cite{Lawden53}, this can only be the optimal strategy if $| \Delta \vec{v}\,| \leq V$. (The range of possible values of $| \Delta \vec{v}\,|$ is a characteristic of the rocket, dictated by its fuel supply and ejection speed). This is easy to understand in the case when the central body is a point mass: indeed, if the rocket is capable of $| \Delta \vec{v}\,| > V$, then it can instead use part of its fuel to reduce its velocity to an arbitrarily small value, entering an elliptical orbit whose pericenter is arbitrarily close to $M$; its velocity at the pericenter is then arbitrarily large, and a second tangential acceleration results in an arbitrarily large energy gain.

Noticing that the rocket's angular momentum (per unit mass) after the instantaneous acceleration is $L=r(V+| \Delta \vec{v}\,|)$, we can express Lawden's bound as 
\begin{equation}\label{Lawden}
\frac{L^2}{r^2} \leq \frac{4M}{r}.
\end{equation}
In this paper, we use the theory of optimal rocket trajectories in general relativity, recently developed in \cite{HN11}, to obtain the general relativistic version of \eqref{Lawden}:
\begin{equation}\label{RelLawden}
\frac{L^2}{r^2} \leq \frac{\frac{M}r\left( 1 - \frac{3M}r \right)^2}{\left( 1 - \frac{3M}r \right)^3 - \frac34 \left( 1 - \frac{6M}r \right)\left( 1 - \frac{2M}r \right)^2} \;\;.
\end{equation}
This bound assumes that the rocket is initially on a {\em stable} circular orbit ($r > 6M$). Note that \eqref{RelLawden} reduces to \eqref{Lawden} in the nonrelativistic limit $\frac{M}r \ll 1$; the two formulas differ significantly for circular orbits close to a compact central object, like a neutron star or a black hole. Due to the more complicated structure of general relativity, there is no simple interpretation of \eqref{RelLawden}, even in the case when the central body is a black hole (the relativistic analogue of a point mass).

For comparison's sake, we must have
\begin{equation}\label{escape}
\frac{L^2}{r^2} \geq \frac{\frac{2M}r}{\left( 1 - \frac{2M}r \right)}
\end{equation}
to escape the gravitational field. Since the right-hand sides of \eqref{RelLawden} and \eqref{escape} become equal for $r=(5+\sqrt{7})M$, we see that for $r<(5+\sqrt{7})M$ the optimal manoeuvre to escape the gravitational field is {\em never} a single tangential instantaneous acceleration.

The structure of the paper is as follows. In Section~\ref{section1} we summarize the theory of optimal rocket trajectories in general relativity, for the convenience of the reader. In Section~\ref{section2} we obtain the endpoint conditions for the problem of maximizing the rocket's energy in a stationary spacetime. Finally, the bound \eqref{RelLawden} is deduced in Section~\ref{section3}.
%
%
\section{The rocket problem in general relativity}\label{section1}
In \cite{HN11}, the classical Newtonian theory of optimal rocket trajectories, developed in \cite{Lawden63}, was generalized to the general relativity setting. There it was shown that optimal trajectories are continuous, sectionally smooth timelike curves, obtained by piecing together free-fall (geodesic) and accelerated arcs, possibly with instantaneous (Dirac delta) accelerations at the junction points. They must satisfy the differential equations
\begin{equation} \label{diffeqns}
\begin{cases}
\displaystyle \nabla_U U^\mu = \frac{a}{\rho} P^\mu \\
\nabla_U P^\mu = - q^\mu + \rho a U^\mu \\
\nabla_U q_\mu = R_{\mu\alpha\beta\gamma}U^\alpha P^\beta U^\gamma
\end{cases}
\end{equation}
where $U$ is the four-velocity, $a$ is the magnitude of the proper acceleration, $P$ is an auxiliary vector field (called the {\bf primer}), $\rho$ is the magnitude of the primer, and $R_{\alpha\beta\mu\nu}$ are the components of the Riemann curvature tensor. All quantities above are continuous at a junction point except possibly $U$ and $P$, which change by a boost with positive parameter in their common plane if there is an instantaneous acceleration. The magnitude of $\rho$ the primer, however, is $C^2$ at the junction points, and satisfies $\rho \leq \rho_0$ on free-fall arcs and $\rho=\rho_0$ on accelerated arcs and instantaneous accelerations (for some constant $\rho_0>0$). Moreover, the equations above admit the first integrals 
\begin{equation} \label{firstintegrals}
P_\mu U^\mu = q_\mu U^\mu = 0.
\end{equation}

The primer $P$ and the vector $q$ are related to the momenta $m_{\mu}$ and $p_\mu$ conjugated to the variables $x^\mu$ and $U^\mu$ by
\begin{equation} \label{mulitplierP}
P_\mu = p_\mu + (p_\alpha U^\alpha) U_\mu
\end{equation} 
and
\begin{equation} \label{mulitplierq}
q_\mu = m_\mu - \Gamma^\alpha_{\beta\mu} p_\alpha U^\beta,
\end{equation}
where $\Gamma^\alpha_{\beta\mu}$ are the Christoffel symbols. If the final values $x^\mu_1$ and $U^\mu_1$ of the variables $x^\mu$ and $U^\mu$ are not specified, then the final values ${(m_\mu)}_1$ and ${(p_\mu)}_1$ of the momenta must satisfy
\begin{equation} \label{mfinal}
{(m_\mu)}_1 = - \frac{\partial J}{\partial x^\mu_1}
\end{equation}
and
\begin{equation} \label{pfinal}
{(p_\mu)}_1 = - \frac{\partial J}{\partial U^\mu_1},
\end{equation}
where $J=J(x^0_1, \ldots, x^d_1, U^0_1, \ldots, U^d_1)$ is the quantity to be minimized. We shall make use of these formulas in the next section.
%
%
\section{Maximizing the final energy}\label{section2}
To be able to speak of energy we consider a stationary $(d+1)$-dimensional spacetime with future-pointing timelike Killing vector field $T$. The rocket's energy per unit mass is then the quantity $E=-T_\mu U^\mu$, and in order to maximize its final value we minimize
\[
J(x^0_1, \ldots, x^d_1, U^0_1, \ldots, U^d_1) = T_\mu(x^0_1, \ldots, x^d_1) U^\mu_1.
\]
If we do not fix $x^\mu_1$ and $U^\mu_1$ then \eqref{mfinal} and \eqref{pfinal} imply
\[
{(m_\mu)}_1 = - \frac{\partial J}{\partial x^\mu_1} = - {(\partial_\mu T_\nu)}_1 U^\nu_1
\]
and
\[
{(p_\mu)}_1 = - \frac{\partial J}{\partial U^\mu_1} = - {(T_\mu)}_1,
\]
or, in view of \eqref{mulitplierP} and \eqref{mulitplierq},
\begin{equation}\label{endpointP}
{(P_\mu)}_1 = {( - T_\mu- (T_\alpha U^\alpha) U_\mu)}_1
\end{equation}
and
\begin{align}
\label{endpointq} {(q_\mu)}_1 & = {(- (\partial_\mu T_\nu) U^\nu + \Gamma^\alpha_{\beta\mu} T_\alpha U^\beta)}_1 \\ 
\nonumber & = {( - (\nabla_\mu T_\nu) U^\nu )}_1 =  {( (\nabla_\nu T_\mu) U^\nu )}_1 = {(\nabla_U T_\mu)}_1,
\end{align}
where we used the Killing equation $\nabla_\mu T_\nu + \nabla_\nu T_\mu = 0$.

Since energy is conserved along geodesics, we can always assume that the optimal trajectory ends on a free-fall arc. Along such arcs we have $a=0$, and so \eqref{diffeqns} reduce to the geodesic and Jacobi (geodesic deviation) equations:
\[
\begin{cases}
\displaystyle \nabla_U U^\mu = 0 \\
\nabla_U P^\mu = - q^\mu \\
\nabla_U q_\mu = R_{\mu\alpha\beta\gamma}U^\alpha P^\beta U^\gamma
\end{cases}
\]
We can obtain a solution to these equations by choosing
\begin{equation} \label{finalarc}
\begin{cases}
P_\mu = - T_\mu- (T_\alpha U^\alpha) U_\mu \\
q_\mu = \nabla_U T_\mu
\end{cases}
\end{equation}
as $T_\alpha U^\alpha$ is constant along the geodesic and both $T$ and $U$ are trivially Jacobi fields. Since these choices also satisfy the endpoint conditions \eqref{endpointP} and \eqref{endpointq}, they must be {\em the} solution in the final free fall arc.
%
%
\section{Optimality condition}\label{section3}
We now focus on the problem of escaping a stable circular orbit in the Schwarzschild spacetime. For simplicity, we restrict ourselves to motions on the $(2+1)$-dimensional totally geodesic submanifold corresponding to the equatorial plane, whose metric in the standard Schwarzschild coordinates $(t,r,\varphi)$ is
\[
ds^2 = - \left( 1 - \frac{2M}r \right) dt^2 +  \left( 1 - \frac{2M}r \right)^{-1} dr^2 + r^2 d \varphi^2.
\]
The geodesic equations in this coordinate system can be written as
\begin{equation}\label{geodesic}
\begin{cases}
\displaystyle \ddot{t} + \frac{2M}{r^2}\left( 1 - \frac{2M}r \right)^{-1} \dot{t} \dot{r} = 0 \Leftrightarrow \left( 1 - \frac{2M}r \right) \dot{t} = E\\ \\
\displaystyle \ddot{r} + \frac{M}{r^2}\left( 1 - \frac{2M}r \right) \dot{t}^2 - \frac{M}{r^2}\left( 1 - \frac{2M}r \right)^{-1} \dot{r}^2 - r \left( 1 - \frac{2M}r \right) \dot{\varphi}^2 = 0 \\ \\
\displaystyle \ddot{\varphi} + \frac2r \dot{r} \dot{\varphi} = 0 \Leftrightarrow r^2 \dot{\varphi} = L
\end{cases}
\end{equation}
where $E$ and $L$ are integration constants (energy and angular momentum per unit mass). Using the normalization condition
\begin{equation}\label{normalization}
- \left( 1 - \frac{2M}r \right) \dot{t}^2 +  \left( 1 - \frac{2M}r \right)^{-1} \dot{r}^2 + r^2 \dot{\varphi}^2 = -1
\end{equation}
we find for a circular orbit traversed in the positive sense ($\dot{r}=\ddot{r}=0$, $\dot{\varphi}>0$)
\begin{equation}\label{circular}
\begin{cases}
\displaystyle E = \frac{1 - \frac{2M}r}{\sqrt{1 - \frac{3M}r}} \\ \\
\displaystyle L = \sqrt{\frac{Mr}{1 - \frac{3M}r}}
\end{cases}
\end{equation}
The Jacobi equation for such a circular orbit was studied in \cite{HN11}. There it was found that the radial and tangential components of the primer satisfy
\begin{equation} \label{ddot{P^r}}
\begin{cases}
\displaystyle \ddot{P}^r + \left( 1 - \frac{6M}r \right) \frac{L^2}{r^4} P^r = 0 \\ \\
\displaystyle \dot{P}^\varphi = - \frac{2L}{r^3} P^r
\end{cases}
\end{equation}
and so, for {\em stable} circular orbits ($r > 6M$) we have
\[
\begin{cases}
\displaystyle P^r = A \sin(\omega \tau) \\ \\
\displaystyle P^\varphi = \frac{2L}{r^3} \frac{A}{\omega} \cos(\omega \tau) + B
\end{cases}
\]
where $A$ and $B$ are integration constants, $\tau$ is the proper time (chosen such that $P^r=0$ for $\tau=0$), and 
\begin{equation} \label{omega}
\omega = \frac{L}{r^2} \sqrt{1 - \frac{6M}r}.
\end{equation}
It was also shown in \cite{HN11} that the magnitude $\rho$ of the primer satisfies
\begin{align*}
\rho^2 & = \left( 1 - \frac{2M}r \right)^{-1} \left( \left(P^r\right)^2 + \left( 1 - \frac{3M}r \right) r^2 \left(P^\varphi\right)^2 \right).
\end{align*}
Therefore we find for $\tau=0$
\begin{equation} \label{rhosquared}
\rho^2 = \left( 1 - \frac{2M}r \right)^{-1}  \left( 1 - \frac{3M}r \right) r^2 \left(\frac{2L}{r^3} \frac{A}{\omega} + B\right)^2,
\end{equation}
$\dot{\rho}=0$ and
\begin{equation} \label{rhoddotrho}
\rho \ddot{\rho} = \left( 1 - \frac{2M}r \right)^{-1}\left( A^2 \omega^2 - \left( 1 - \frac{3M}r \right) \frac{2LA\omega}{r} \left(\frac{2L}{r^3} \frac{A}{\omega} + B\right) \right).
\end{equation}
From \eqref{circular}, \eqref{omega} and \eqref{rhosquared}, and assuming $P^\varphi>0$ for $\tau=0$, we find
\[
\rho = \left( 1 - \frac{2M}r \right)^{-\frac12}  \left( 1 - \frac{3M}r \right)^{\frac12} \left(\frac{2A}{\sqrt{1 - \frac{6M}r}} + Br \right),
\]
and so
\begin{equation} \label{A}
A = \frac12 \left( 1 - \frac{2M}r \right)^{\frac12}  \left( 1 - \frac{3M}r \right)^{-\frac12} \left( 1 - \frac{6M}r \right)^{\frac12} (\rho - D),
\end{equation}
where
\[
D = \left( 1 - \frac{2M}r \right)^{-\frac12}  \left( 1 - \frac{3M}r \right)^{\frac12} rB. 
\]
Using \eqref{rhosquared} and \eqref{A}, we can write \eqref{rhoddotrho} as
\begin{align} \label{rhoddotrho0.5}
\rho \ddot{\rho} & = \left( 1 - \frac{2M}r \right)^{-1} A \omega \left( A \omega - \left( 1 - \frac{3M}r \right)^{\frac12}  \left( 1 - \frac{2M}r \right)^{\frac12} \frac{2L}{r^2} \rho \right) \\
\nonumber & = \frac{\frac{M}{r^3}\left( 1 - \frac{6M}r \right)}{\left( 1 - \frac{3M}r \right)} (\rho - D) \left[ \frac{\left( 1 - \frac{6M}r \right)}{4 \left( 1 - \frac{3M}r \right)}(\rho - D) - \rho \right].
\end{align}

Now consider the manoeuvre in which the rocket escapes the circular orbit through a tangent instantaneous acceleration in the positive sense. Since for such a manoeuvre to be optimal we must have $P^r=0$ and $P^\varphi>0$ (recall that the primer determines the direction of the instantaneous acceleration), we can assume without loss of generality that the instantaneous acceleration takes place at $\tau=0$. For $\tau>0$ we then have, by \eqref{finalarc},
\[
P = - \frac{\partial}{\partial t} + E U,
\]
where $E$ is the energy per unit mass of the final free-fall arc, and so
\begin{equation} \label{rhosquared2}
\rho^2 = P_\mu P^\mu = - \left( 1 - \frac{2M}r \right) - E^2 + 2E^2 = E^2 - 1 + \frac{2M}r.
\end{equation}
Since $\dot{r}=0$ for $\tau = 0$, we have
\begin{equation} \label{rhoddotrho2}
\rho \ddot{\rho} = - \frac{M}{r^2} \ddot{r} = \frac{M^2}{r^4}\left( 1 - \frac{2M}r \right)^{-1} E^2 - \frac{M}{r^5} \left( 1 - \frac{2M}r \right) L^2,
\end{equation}
where we used \eqref{geodesic}. Since by \eqref{normalization} we have
\[
\left( 1 - \frac{2M}r \right)^{-1} E^2 = 1 + \frac{L^2}{r^2},
\]
we can rewrite \eqref{rhosquared2} and \eqref{rhoddotrho2} as
\begin{equation} \label{rhosquared3}
\rho^2 = \left( 1 - \frac{2M}r \right) \frac{L^2}{r^2}
\end{equation}
and
\begin{equation} \label{rhoddotrho3}
\rho \ddot{\rho} = \frac{M}{r^3}\left[ \frac{M}r - \left( 1 - \frac{3M}r \right) \frac{L^2}{r^2} \right].
\end{equation}
Since $\rho$ must be $C^2$, the values of $\rho \ddot{\rho}$ from expressions \eqref{rhoddotrho0.5} and \eqref{rhoddotrho3} must be the same. Equating them and using \eqref{rhosquared3} we obtain
\begin{equation} \label{main}
\frac{\left( 1 - \frac{3M}r \right)}{\left( 1 - \frac{2M}r \right)} \rho^2 + \frac{\left( 1 - \frac{6M}r \right)}{\left( 1 - \frac{3M}r \right)} (\rho - D) \left[ \frac{\left( 1 - \frac{6M}r \right)}{4 \left( 1 - \frac{3M}r \right)}(\rho - D) - \rho \right] = \frac{M}r.
\end{equation}

Notice that $\rho$ must have a local maximum at $\tau=0$, and so we must have $\ddot{\rho}\leq 0$, which by \eqref{rhoddotrho0.5} implies $A \geq 0$, that is, $\rho \geq D$. On the other hand, the maximum must be the global maximum of $\rho$ over the circular orbit, since if one waits for an arbitrary amount of proper time before performing the instantaneous acceleration the trajectory thus obtained must still be optimal. Therefore we must have $B \geq 0$, that is, $D \geq 0$.

Consider the function
\[
f(D) = (\rho - D) \left[ \frac{\left( 1 - \frac{6M}r \right)}{4 \left( 1 - \frac{3M}r \right)}(\rho - D) - \rho \right].
\]
We have
\[
f'(D) = \rho - \frac{\left( 1 - \frac{6M}r \right)}{2 \left( 1 - \frac{3M}r \right)}(\rho - D) 
\]
and so $f'(D)>0$ for $\rho > 0$, $r>6M$ and $D\geq 0$. Consequently $f(D) \geq f(0)$, and so \eqref{main} implies
\[
\frac{\left( 1 - \frac{3M}r \right)}{\left( 1 - \frac{2M}r \right)} \rho^2 + \frac{\left( 1 - \frac{6M}r \right)}{\left( 1 - \frac{3M}r \right)} \rho^2 \left[ \frac{\left( 1 - \frac{6M}r \right)}{4 \left( 1 - \frac{3M}r \right)} - 1 \right] \leq \frac{M}r,
\]
which, using \eqref{rhosquared3}, is easily seen to be equivalent to
\[
\frac{L^2}{r^2} \leq \frac{\frac{M}r\left( 1 - \frac{3M}r \right)^2}{\left( 1 - \frac{3M}r \right)^3 - \frac34 \left( 1 - \frac{6M}r \right)\left( 1 - \frac{2M}r \right)^2}.
\]
%
%

\end{document}